# Structural and Optical properties of $Zn_{1-x}Mg_xO$ nanocrystals obtained by low temperature method

Manoranjan Ghosh [@], and A.K.Raychaudhuri[*]

*DST Unit for Nanoscience, S.N. Bose National Centre for Basic Sciences*
*Block- JD, Sector-III, Salt Lake, Kolkata - 700 098, INDIA.*

## Abstract

In this paper we report structural and optical properties of Magnesium substituted Zinc Oxide ($Zn_{1-x}Mg_xO$) nanocrystals (~10-12nm) synthesized by low temperature route. In the low temperature synthesis route it was possible to reach x = 0.17 without segregation of Mg rich phase. The exact chemical composition has been established by quantitative analysis. Rietveld analysis of the XRD data confirms the Wurzite structure and a continuous compaction of the lattice (in particular the c-axis parameter) as x increases. There is an enhancement of the strain in the lattice as the Mg is substituted. The bandgap also gets enhanced as x is increased and reaches a value of 4eV for x = 0.17. From the TEM and the XRD data it has been concluded that when there is a phase segregation for x > 0.17, there is a shell of $Mg(OH)_2$ on the ZnO. The absorption also shows persistence of the excitoinc absorption on Mg substitution. The nanocrystals show near band edge photo luminescence (PL) at room temperature which shows blue shift on Mg incorporation. In addition to the near band edge emission the ZnO and $Zn_{1-x}Mg_xO$ alloy nanocrystals show considerable emission in the blue-green region at wavelength of ~550 nm. We find that the relative intensity of the green emission increases with the Mg concentration for very low x (upto x = 0.05) and on further increase of the Mg concentration there is a sharp decrease of relative intensity of the green emission eventually leading to a complete quenching of blue emission. It is concluded that due to phase segregation (for x ≥ 0.20), the formation of the shell of $Mg(OH)_2$ on the ZnO leads to quenching of the green emission .However, this shell formation does not have much effect on the near band edge PL.

@ email: mghosh@bose.res.in

* email: arup@bose.res.in, also at the Department of Physics, Indian Institute of Science, Bangalore –560012, INDIA

## 1. Introduction

ZnO is a wide band gap semiconductor. It has a Wurzite structure with lattice constants of a = b = 3.25 Å and c = 5.207 Å. It has a direct bandgap of 3.3eV and the fundamental absorption even at room temperature shows excitonic contribution due to its large excitonic binding energy (~ 60meV)[1,2]. One of the interesting features of ZnO is the possibility to tune its band gap by substituting (alloying) bivalent metals like Cd and Mg in place of Zn to its band gap. While Cd is known to reduce the band gap[3], Mg substitution leads to enhancement of band gap[4]. This raises the possibility of band gap engineered heterostructure for optoelectronics applications in the UV range[5].

Synthesis of ZnO for optoelectronics applications are generally done through vapor phase deposition with or without catalyst. Routes that have been used to prepare ZnO and $Zn_{1-x}Mg_xO$ alloys are Molecular Beam Epitaxy (MBE)[6], Vapor liquid solid phase growth[4] Pulsed Laser Deposition (PLD)[7] and Metallo-organic Vapor Phase Epitaxial growth (MOVPE)[8]. Materials obtained are mostly films or rods. In addition, a very low temperature (≤ 100 °C) solution route has been used to fabricate aligned nanorods of ZnO[9]. In a recent paper Mg substituted ZnO ($Zn_{1-x}Mg_xO$) nanorod synthesized by hydrothermal method (100°C) has been reported[10]. The advantage of the solution route is that it can give high quality crystalline materials through an energy efficient route that allows wide area growth (e.g., 4" Si wafer). Our principal motivation is to explore whether such a low temperature route will allow us to grow well-characterized Mg substituted nanocrystals of ZnO as an optical material. In the solution route for material synthesis the mixing occurs at the atomic level and the material made by this method is close to the thermodynamic equilibrium phase.

One of the important issues that limits synthesis of MgO alloyed ZnO is the fact that the thermodynamic solubility limit of MgO in ZnO is ≈ 4 at. %[11]. The issue of Mg incorporation in ZnO in excess of 4 at% without phase segregation is an important and topical research problem. It has been claimed that using the vapor phase route such as pused laser deposition it is possible to incorporate Mg up to ~ 40 at.%[4]. For MOVPE grown films the Mg incorporation has been reported to be ~49 at%.[8]. For ZnO grown solution route x~ 0.25 has been reported to be achieved[10]. In this paper we would like to explore a very specific question- to what extent one can incorporate Mg into ZnO nanocrytals using low temperature solution route (growth temperature < 70°C) and the resulting quality of the material as characterized by structural and optical means. We show that such a low temperature route can be used to make the alloy $Zn_{1-x}Mg_xO$ with x as large as 0.17. Above this composition the MgO segregates. Up to this composition the alloy retains a Wurzite structure and the lattice constants changes smoothly as x is varied albeit by a small amount. The phase segregation for x > 0.20 has been established by XRD as well as TEM studies. However, even with such low incorporation of Mg there are important and topical changes in the optical properties of the material. The alloy nanocrystals show an enhanced band gap that reaches ≈ 4.0 eV for x ≈ 0.17 and an appreciable near band edge Photoluminescence (PL) response when excited with UV radiation. The near band edge UV emission occurs at room temperature and the emission shows blue shift as x is increased. The choices of nanocrystals rather than nanorods for this work are two fold. First, working with good quality nanocrystals allows us to use the Rietveld analysis[12] for the x-ray diffraction to get a more quantitative evaluation of lattice constants, which is not possible with aligned rods. Second, ZnO nanocrystals showing good optical properties are basic building blocks for all applications and can often be used directly for many optoelectronic applications. It is like a quantum dot operating in the UV.

While investigating the effect of Mg substitution on the band gap one needs caution. This is because if the particle size is smaller than 5 nm, the effect of quantum confinement plays a significant role and the band gap becomes a function of the size[13]. While making the alloy, if a strict control of the size is not achieved then two effects (one from quantum confinement, and other due to Mg substitution) will make comparable contribution and one may not be able to separate out the two effects. In this work we used nanocrystals in the size range 10-12 nm. In this range the quantum effects make negligible contribution. As a result we can cleanly isolate the effect of Mg substitution on the optical properties.

## 2. Experimental

ZnO nanoparticles were synthesized using acetate route. 0.03 M NaOH solution in ethanol is slowly added in 0.01 M solution of $Zn(CH_3COO)_2$, $2H_2O$ in ethanol kept at 65° C. The final solution was stirred and heated at 65 °C for two hours. This method allows precipitation of ZnO nanoparticles and avoids precipitation of hydroxides if the temperature is more than 60 °C. For the purpose of making alloy, we have added $Mg(CH_3COO)_2$, $4H_2O$ with the $Zn(CH_3COO)_2$, $2H_2O$ solution and same procedure as described above have been followed. To vary the doping concentration, the relative concentration of $Mg(CH_3COO)_2$, $4H_2O$ and $Zn(CH_3COO)_2$, $2H_2O$ in the solution were varied. The temperature at which reaction occurs can be varied from 60 °C to 75 °C. To eliminate the large particle, the samples in solution phase were centrifuged for 2 minute at 2000 rpm. The dispersion containing the smaller particles were again centrifuged for 6 minute at 9000 rpm. The precipitates were collected, washed by water and centrifuged again at 9000 rpm. The final precipitates were collected by dispersing them in ethanol. All the optical measurements have been performed in solution phase taking the dispersed sample within a cuvette. For XRD investigation, washed residues were collected and dried within a hot air oven at temperature 120° C.

The chemical compositions (x) were confirmed by Inductively Coupled Plasma Atomic Emission Spectroscopy (ICP-AES)) and Energy Dispersive X-ray analysis. (We note that in general, most past studies (since they were mostly done on film) could not provide an independent quantitative analysis of exact Mg content after alloying ).

The nanoparticles were imaged using TEM and their lattice images were taken with a JEOL HR-TEM (at 200 KeV). The crystal structure of the ZnO and $Zn_{1-x}Mg_xO$ alloy nanocrystals were studied by means of Rietveld analysis of the X-ray diffraction data. Combination of XRD and the TEM studies helped us to analyze what happens when Mg is progressively alloyed in ZnO. Room temperature optical properties were studied by UV–visible spectrophotometer (Shimadzu UV-2450) and photoluminescence spectrometer (JOBIN YVON – FluoroMax–3) where a Xenon arc-lamp has been used as illumination source.

## 3. Results and discussion

3.1 ICP-AES AND TEM ANALYSIS: The basic characterization of the nanoparticles were done through Inductive Coupled Plasma Atomic Emission Spectroscopy (ICP-AES) and TEM in addition to XRD. The result of the XRD analysis are given in the subsequent sub-section. We determine the exact chemical composition of the particles using the ICP-AES technique which show that the maximum composition of Mg that we could get into the nanocrystals correspond to $x \approx 0.17$. The analysis of the chemical composition by analytical quantitative technique like the ICP-AES is important because it establishes exactly how much Mg can be incorporated into ZnO. This is larger than the equilibrium composition ($x \sim 0.04$) but much less than what has been reported for vapor phase methods[4]. In figure.1 we show the value of exact 'x' as determined through ICP-AES analysis vs the molar fraction of $Mg(CH_3COO)_2$, $4H_2O$ added in the $Zn(CH_3COO)_2$, $2H_2O$ solution. For samples greater than x = 0.17 the resulting nanocrystals is not single phase as we will show form the XRD and the TEM data. They show phase separation between two phases viz. wurtzite ZnO and brucite $Mg(OH)_2$ (this will be seen more clearly from the XRD investigation discussed later). Sample with x = 0.52 show a thin layer of $Mg(OH)_2$ on the $Zn_{1-x}Mg_xO$ nanoparticles forming a core-shell structure.

Recently it has been established that the doping efficiency is determined by the binding energy of dopant atoms on the individual nanocrystal surface[14]. The binding energy for wurtzite nanocrystal is three times less than that of zinc blend or rocksalt structure on (001) faces. Therefore the low concentration of incorporated Mg is likely to be a consequence of lower doping efficiency of the wurtzite ZnO nanocrystals surfaces.

TEM images of the ZnO (undoped) nanocrystals with a uniform size distribution (average diameter~ 10 nm) are shown in figure 2. Figure 2(a) shows the TEM image of a collection of the particles. The lattice fringes and selected area electron diffraction (SAED) pattern of a single particle is shown in fig. 2(b) and 2(c) respectively. The SAED confirms the hexagonal symmetry of the particles. The distance between the parallel lattice fringes of the undoped ZnO particle is 2.62 Å. The particles obtained by this method are well defined crystals. In figure 3(a) we show a typical HR-TEM image of an alloy ($Zn_{1-x}Mg_xO$) nanoparticle with x=0.17. The distance between the parallel lattice planes is very similar to that of the undoped crystals (~2.60Å). Figure 3.(b) shows the SAED pattern of the same which confirms that the hexagonal symmetry of the ZnO nanocrystals are preserved on Mg substitution. The TEM images clearly establish that the Mg incorporation does not lead to any change in the crystal symmetry and the crystal structure upto about x ≈ .17. This is an important observation because MgO has a rock salt structure and for higher concentration of Mg one would expect segregation of phases.

In the following we establish through the TEM study (which will be later elaborated by XRD ) that on further addition of Mg there is no incorporation or alloying of Mg into the lattice. Instead there is a phase segregation. We observe (presented below) that for higher Mg incorporation there is a clear appearance of XRD peaks due to $Mg(OH)_2$ brucite structure. Figure. 4 shows the TEM images of the same sample grown with condition of 55 mole % $Mg(CH_3COO)_2$, $4H_2O$ in the solution (x = 0.52) which clearly shows a thin layer of $Mg(OH)_2$ on the $Zn_{1-x}Mg_xO$ core. Upper part of the figure represent a single particle which shows a thin ring around the core. Lower part shows the HRTEM image taken at the junction of the thin layer and the core which clearly establishes the existence of two phases on the two sides of the junction. The left one shows uniformly spaced lattice fringes and two spots on the SAED pattern (left most part, indicated by an arrow) which confirms the existence of single phase. The phase was identified as $Mg(OH)_2$ by the XRD data as it shows two distinct peaks (corresponding to two spots in the SAED pattern) due to (100) and (110) planes of brucite $Mg(OH)_2$. On the right side of the junction, lattice fringes are overlapped and this indicates mixture of two phases (wurtzite $Zn_{1-x}Mg_xO$ with $Mg(OH)_2$ as cover layer). This fact is also clear from the SAED pattern ( right most figure indicated by an arrow) which shows diffused pattern, again indicating a phase mixture. From the XRD data, the core structure of this sample can be easily identifiable as wurtzite $Zn_{1-x}Mg_xO$.

To summarize this section we find that Mg can be incorporated upto x ≈ 0.17 in the alloy $Zn_{1-x}Mg_xO$ nanoparticles by low temperature solution method without phase segregation and retaining the Wurzite structure of the parent ZnO system. The resulting nanoparticles (diameter ~ 10-12 nm) are single crystals. Further addition leads to phase segregation that creates a core of the alloy $Zn_{1-x}Mg_xO$ and a shell of the $Mg(OH)_2$. (Note: The EDX taken in the TEM also allows one to evaluate the Mg concentration in the nanoparticles.

A comparison with ICP-AES measurements shows that the concentration determined by EDX alone overestimates x severely. We have used the ICP-AES results because the method is more standardized with available standards.)

3.2 X-RAY DIFFRACTION AND ANALYSIS OF STRUCTURAL DATA: One of the important issue in the alloy $Zn_{1-x}Mg_xO$ formation is the change in the lattice constant when Mg is incorporated. $Mg^{2+}$ and $Zn^{2+}$ have similar ionic radii - 0.57Å and 0.60 Å respectively[15]. The alloy formation is thus expected to give a small change in the lattice constants due to the smaller size (~5%) of the Mg ions. Although Zn and Mg have very comparable ionic radii, the crystal structures of MgO and ZnO are different and it is expected that at certain concentration of Mg the simple substitution has to terminate. As shown in figure 5(a), the observed powder data for $Zn_{1-x}Mg_xO$ can be indexed to wurtzite structure upto x =0.17. This indicates that single phase $Zn_{1-x}Mg_xO$ nanoparticles form with Mg incorporation. However, for higher Mg content ( x >0.17) additional peaks appear in the XRD data. These two peaks are due to (100) and (110) reflections of $Mg(OH)_2$ as indicated in figure 5(a). This indicates phase separation between ZnO and $Mg(OH)_2$. We have seen before in the TEM image (figure 4) that this sample shows a core-shell structure of a thin layer of $Mg(OH)_2$ on the $Zn_{1-x}Mg_xO$ core .

The crystal structure were investigated by analyzing the XRD data ( upto x = 0.17 ) by full line profile fitting using Rietveld method .The fit of the data have been shown as a solid line in figure 5(a). In the least square refinement peak shape was assumed to be Pseudo-voigt. An example of the residue of the fitting is shown in the figure for x = 0 data. The profile fitting to determine the lattice constants assumed that the Mg is on substitutional sites and it occupies these sites randomly. One obtains good fitting with low residue with this assumption. We can thus conclude that the alloy formation does occur with Mg randomly substituting for Zn. The HR-TEM lattice images also corroborates this.

As shown in figure 5(a), the peaks in the powder XRD of the nanocrystals shift to higher angle as x is increased, indicating decrease in the lattice constants of the ZnO due to incorporation of Mg. The lattice constants $(a,b,c)$ and the cell volume $V$ $(=0.866a^2c$ for hexagonal lattice) were determined from the full profile fitting of the observed data. As plotted in figure 5(b) the lattice constants as well as the cell volume decrease monotonously as x is increased. We observed 0.3 % change in the c-axis length which is smaller than the approximate decrease in the value of the c-axis lattice constant ($\approx 0.6\%$ for x=0.2) observed in MOVPE grown $Zn_{1-x}Mg_xO$ alloy films[8]. Thus, the lattice constant depends on the method of growth. The Rietveld analysis shows that the lattice constant $a(=b)$ changes by even a smaller amount ~ 0.15% for the same Mg concentration. The XRD studies show that the main effect of Mg substitution is compaction along the c-axis which in turn leads to a compaction of the unit cell volume.

The XRD data were also used to find the average particle size and strain using Williamson-Hall (W-H) analysis using simplified integral breadth method[16]. The Integral Breadth(β), defined as the ratio between peak area and the intensity maxima for both size and strain broadened profile is given by[16]

$$\beta^* = 1/D + 2.\varepsilon.s \qquad \text{(in sin}\theta \text{ scale)} \qquad (1)$$

where $\beta^* = \beta\cos\theta/\lambda$ , $\varepsilon$ is strain and  s = $2\sin\theta/\lambda$.
The average particle diameter was obtained form this analysis and this match with that found from the TEM. W-H analysis also allows us to find the average strain from the slope of the linear fit of the data points. The strain values calculated by this method were plotted with the Mg content 'x' in fig.6. For low x , the strain is within the range $\varepsilon \approx 1\text{-}5 \times 10^{-3}$ which is similar to that seen in undoped (x=0) nanocrystals. Although there is a gradual increase in $\varepsilon$ as x increases, for $x \geq 0.17$ there is a sudden increase in the strain value by a large amount. Interestingly this is the value of x for which the phase segregation takes place. The XRD data (along with the TEM data) provides clear evidence for the phase segregation. It therefore appears that the phase segregation is driven by the built in strain on Mg incorporation.

To summarize, the extensive analysis the XRD data shows that Mg can be incorporated in the ZnO forming the random substitutional alloy $Zn_{1-x}Mg_xO$ for $x \leq 0.17$. It is important to establish that the material thus formed has the Mg in the substitutional sites. The substitution leads to a compaction of the c-axis lattice constant. However, for higher values of x there is a phase segregation, which is corroborated by TEM results presented in previous subsection.

3.3 OPTICAL PROPERTIES:

UV ABSORPTION: The room temperature UV-visible absorption spectra of ZnO and the $Zn_{1-x}Mg_xO$ alloy nanocrystals were taken by dispersing them in ethanol. As shown in the figure 7(a), a sharp absorption peak appears in the absorption curve due to the large excitons followed by the fundamental absorption. The sharp peak which has been identified as due to excitons and is visible even at room temperatures due to the large exitonic binding energy of ZnO (~ 60meV). The absorption curve shifts to higher energy in the alloy nanocrytals. In the same graph we show the absorption spectra taken at room temperature in a single crystal of ZnO (we refer to this as bulk). Compared to the Bulk the fundamental absorption edge is shifted by about 0.25eV in the nanocrytal (with average diameter ~ 10nm ). We observe a large and gradual blue shift of the absorption edge on Mg substitution. This is shown in figure 7(b) where we plot the shift of the absorption edge energy ($\Delta E_g$) as a function of x for the $Zn_{1-x}Mg_xO$ alloy nanocrystals. Interestingly the $E_g$ shifts to as high as 4eV on Mg substitution. (As pointed out before it is not possible to have a tight control on the particle size on Mg substitution. The particle sizes may vary by about 2nm on the average. However, for an average size of 10nm, such a size variation (~2nm) can only produce a change in $E_g$ of about 0.05 eV [13]. This is much smaller compared to the change in $E_g$ seen on Mg substitution. For the phase segregated sample (curve 5 in fig 7(a)), no more shift in the band gap value has been observed. Also the excitonic peak disappear from the absorption spectrum.

NEAR BAND EDGE EMISSION: One of the exciting thing about ZnO is its ability to show near band edge Photoluminescence (PL) even at room temperatures when irradiated by radiation with energies somewhat higher than the $E_g$. At room temperature in addition to the near band edge (NBE) emission a lower energy emission is also seen. The deep level PL is a matter of debate and is now established to be associated with defects. In figure 8(a) we show the NBE emission obtained from the ZnO and alloy nanocrytals. The PL spectra when analysed for different peak showed that it contains two main lines at 347 nm and 377 nm and a weak line at 398 nm. All the samples were excited at 230 nm and have one peak located in the UV region of the spectra which is due to the excitonic near-band-gap emission. In $Zn_{1-x}Mg_xO$ there is a clear blue shift of the NBE as x is increased. This is shown in figure 8 (b). For x = 0.17 the blue shift is nearly 12 nm. This is, however, less than the blue shift observed in the band edge from the absorption data (42nm).

The luminescence show a stokes shift to the lower energy side of the absorption edge which is very common in alloy semiconductors[17]. For instance undoped ZnO (x=0) shows PL peak at 377 nm whereas the $E_g$ occurs at 354nm. On Mg substitution the PL peak shifts to 365 nm while $E_g$=312nm for x = 0.17. The blue shift of NBE emission is because of the doping of Mg in ZnO. Since ZnO is a natural n-type material, the height of the Fermi level increases and will be inside the conduction band due to the incorporation of Mg. The states below the Fermi level in the conduction band being full, the absorption edge shifts to the higher energy[18]. The blue shift of the PL peak with x is also monotonous like the shift in $E_g$.

The NBE emission has details which are not seen in the room temperature spectra. It has been established at least in bulk ZnO by PL investigations at low temperature that the NBE emission actually contains a number of lines that arise from free exciton as well as from exciton bound to impurities. The substitution by Mg can lead to redistribution of energy among various absorption lines to shift in the NBE, in addition to the blue shift of the PL main line. Importantly these changes on substitution are substantial and they happen even with small amount of substitution ( x < 0.15).

The NBE UV emission may get a contribution from Raman scattering arising from the phonon modes. To check, whether it has a contribution from the Raman scattering, we varied the excitation frequency and noted the NBE spectra. We find that the line positions mentioned as PL peaks does not shift with the excitation frequency.

VISIBLE EMISSION: In addition to the near band edge emission the ZnO and $Zn_{1-x}Mg_xO$ alloy nanoparticles show considerable emission in the blue-green region at wavelength of ~550 nm. This particular emission occurs in almost all samples of ZnO and its origin has been a topic of much discussion[19]. It is however, accepted that this emission has an extrinsic origin (defect mediated). The extrinsic nature of the 550nm emission can be seen from the fact that it depends on the surface condition and also the details of the preparation. In this paper we have also investigated briefly what happens to this emission on Mg substitution. However, the detailed investigation of this specific issue is beyond the scope of the paper. We find that the effect of Mg substitution is two fold. First there is a nonmonotonous dependence of the intensity of the visible emission on x. Second, there is blue shift of the emission peak on Mg substitution.

The green emission, has a strong and non-monotonous dependence on x. The intensity ratio of the visible to near band edge UV emission increases very rapidly up to x ≤0.05. An example of the blue-green emission for low x is shown in figure 9(a) for x = 0.05. Further increase in the Mg content reduces the intensity ratio between the two lines (figure 9(b) ). For x > 0.2, when the core-shell structure sets in there is complete quenching of the visible emission at 550nm, whereas the yield of the UV emission remains same. It is likely that the $Mg(OH)_2$ layer on the ZnO nanocrystals (revealed by the XRD and the TEM data) actually quenches the visible emission at 550nm. The quenching of the visible emission on formation of the core-shell structure is a new observation and has the possibility of its utilization in controlled optical properties of ZnO nanocrystals. A detailed investigation of the specific issue of the visible emission from Mg substituted ZnO is under way and is beyond the scope of the paper.

Another important effect of Mg substitution is the blue shift in the emission spectra as x increases. As shown in fig. 8(b) the visible emission is significantly blue shifted on Mg substitution, ranging from 550nm (x=0) to 514 nm (x = 0.17). On further substitution the emission intensity comes down as noted before.

CONCLUSION: In summary, $Zn_{1-x}Mg_xO$ nanocrystals (with average size ~10-12nm) are synthesized by using a low temperature solution growth method (growth temperature 60-75° C). The crystal structure analysis (based on the XRD data) and the TEM –diffraction pattern show that the $Zn_{1-x}Mg_xO$ nanocrystals have wurtzite structure as that of ZnO upto

x=0.17. Beyond that segregation of $Mg(OH)_2$ occurs. The composition as checked by the ICP-AES shows that EDX composition analysis (as is generally done in doped ZnO films) may overestimate the actual Mg incorporation. Room temperature spectra reveals that Mg substitution leads to enhancement of the fundamental absorption edge from 3.55eV to 3.99 eV and the near band edge emission from 377 to 365 nm for x as little as 0.17. After phase segregation, the excess $Mg(OH)_2$ forms a layer on the nanocrystals, which quenches the visible emission.

ACKNOWLEDGMENT

The authors want to thank Department of Science and Technology for financial support as Unit for Nanoscience. Technical support by Dr. Supriya Chakraborti for use of the TEM facility at Indian Association for the Cultivation of Science is acknowledged.

# Figure captions:

Figure.1: Exact concentration x of Mg in $Zn_{1-x}Mg_xO$ nanocrystals determined by ICP-AES vs the molar fraction of Mg acetate ($Mg(CH_3COO)_2$, $4H_2O$) in solutions.

Figure. 2.(a): TEM image of undoped ZnO nanocrystals. (b) and (c) show the HRTEM image and the SAED of the same sample respectively.

Figure. 3.(a) and (b) show the HR-TEM image and SAED pattern of the $Zn_{1-x}Mg_xO$ nanocrytals with Mg content x = 0.17.

Fig. 4. TEM image showing core-shell structure of $Zn_{1-x}Mg_xO$ coated with $Mg(OH)_2$ seen in phase separated nanocrystals. Upper part of the figure shows a thin layer around the core. The lower part represents the lattice image at the interface of the two materials which clearly shows two different arrays of lattice fringes separately for $Mg(OH)_2$ (left) and mixture of $Zn_{1-x}Mg_xO$ and $Mg(OH)_2$ (right).

Fig. 5. (a): XRD data taken on $Zn_{1-x}Mg_xO$ nanocrystals. It shows single phase (wurtzite structure, peaks are indexed) upto x = 0.17 . Beyond this level phase separation takes place and distinct peaks due to (100) and (110) planes of $Mg(OH)_2$ (indexed in the figure) are observed. The data upto x = 0.17 are analysed by Rietveld method. Observed data (dotted curve) upto x = 0.17 are fitted well with the calculated data (solid line). (examples of residues are shown for x = 0 at the bottom). Fig. 5 (b): Variation of $c$-axis length and cell volume (extracted from the Rietveld analysis of the data) with the Mg content x.

Fig. 6 Variation of strain as obtained from the XRD data with x. The large rise in the strain in the region of phase separation can be seen in the graph.

Fig. 7. (a): Absorption spectra of the $Zn_{1-x}Mg_xO$ nanocrystals for different Mg content. Curve no 5 is for phase segregated sample. Fig. 7(b): Shift in the direct band gap values ($\Delta E_g$) of the $Zn_{1-x}Mg_xO$ alloy nanocrystals compared to undoped ZnO nanocrystal.

Fig. 8.(a): The near band edge luminescence in $Zn_{1-x}Mg_xO$ alloy nanocrystals. The PL shows blue shift as the Mg content increases. The visible emission also shift to the blue side as the Mg content increases. Fig.8(b): The shift in emission wavelength compared to the undoped sample for both the NBE emissions and green emissions as a function of 'x'.

Fig 9(a): An example of the 550nm emission in doped system with x = 0.05. Fig. 9(b): The variation of the intensity ratio between the visible and the UV emission as Mg content increases. Plot shows that the intensity of the visible emission increases rapidly as the Mg content increases but falls drastically as soon as the phase segregation starts. (see text).

**Figures:**

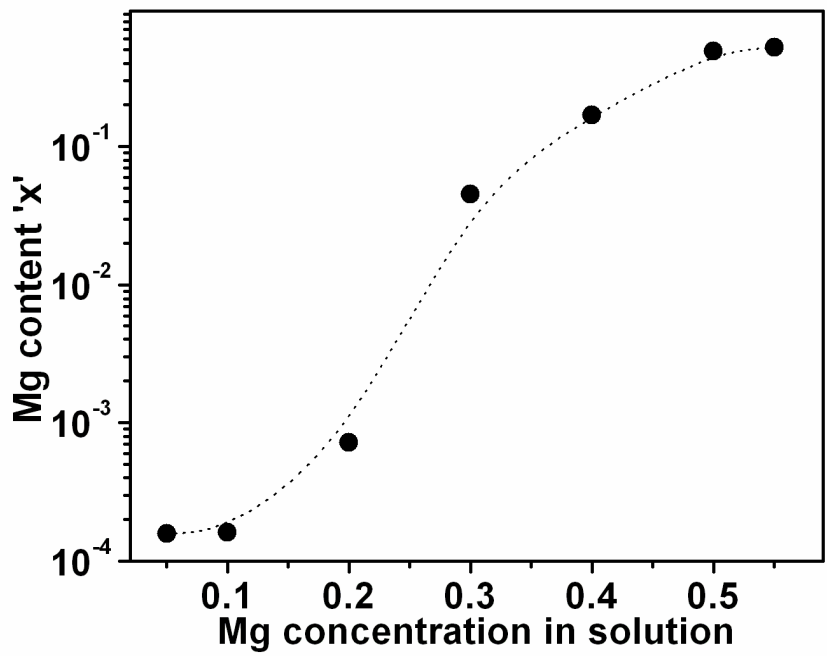

Fig. 1

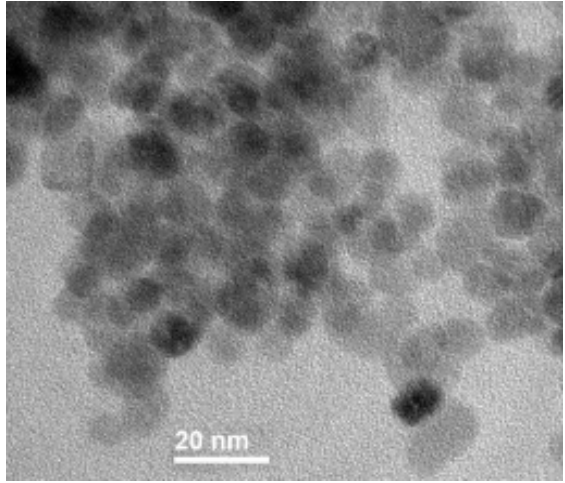

Fig. 2(a)

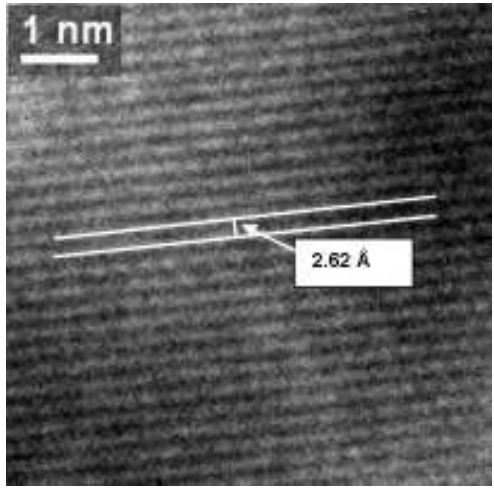

Fig. 2(b)

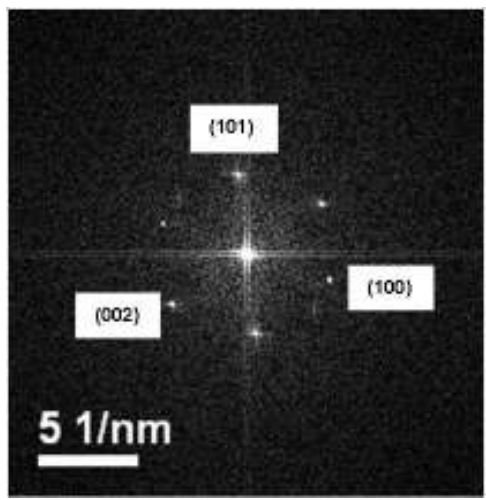

Fig. 2(c)

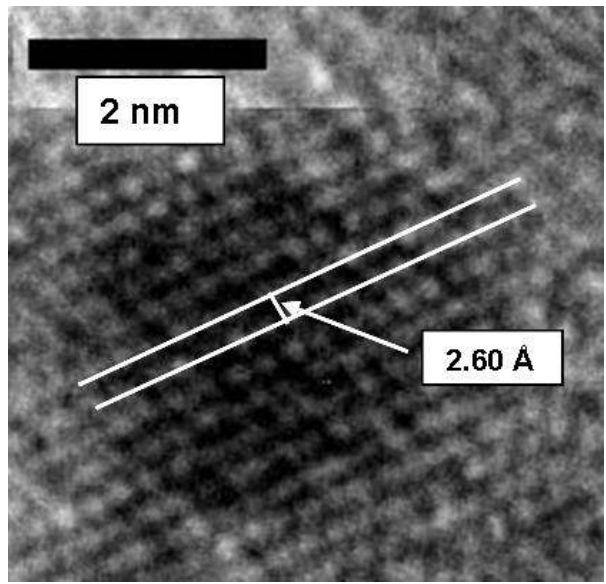

Fig. 3(a)

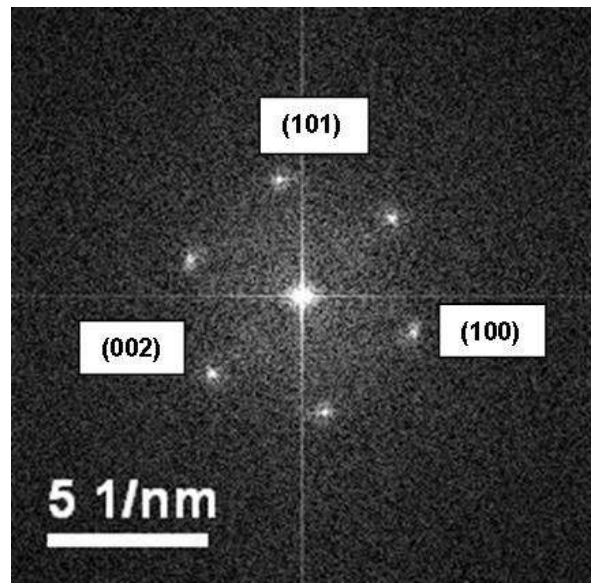

Fig. 3(b)

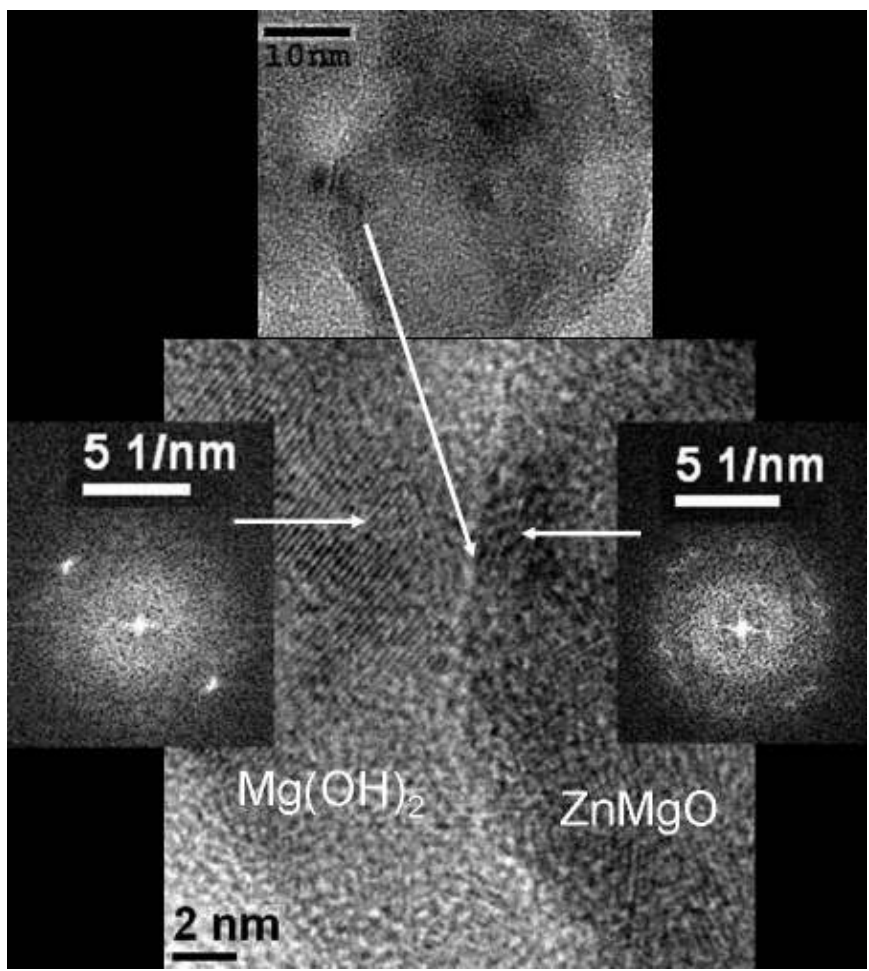

Fig. 4

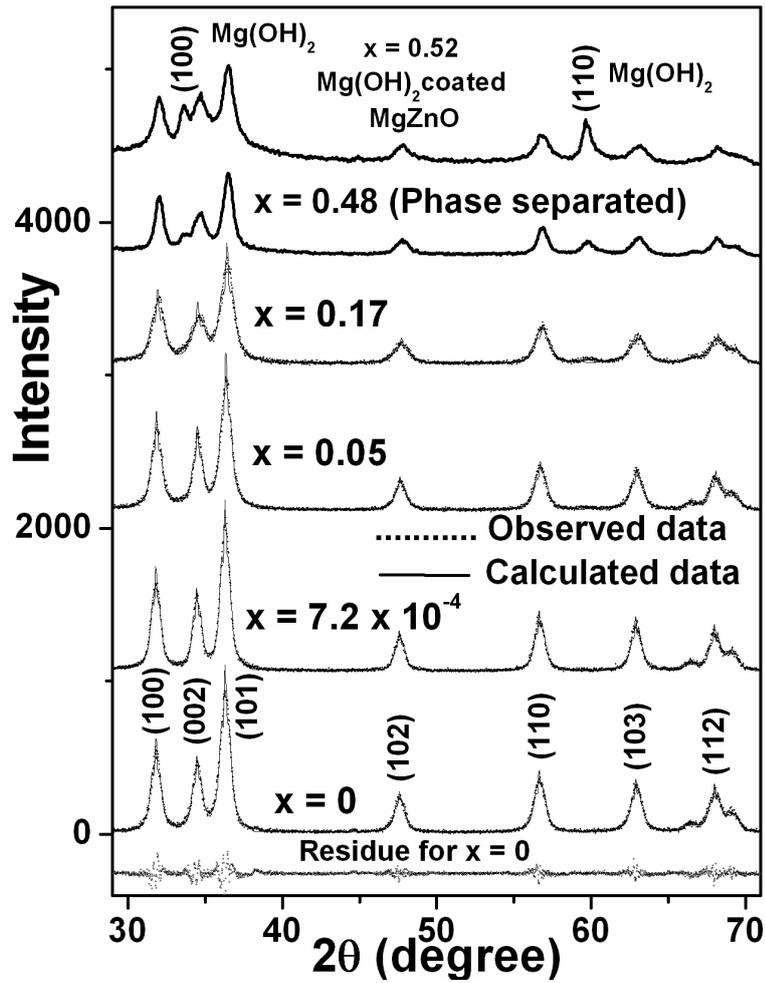

**Fig. 5(a)**

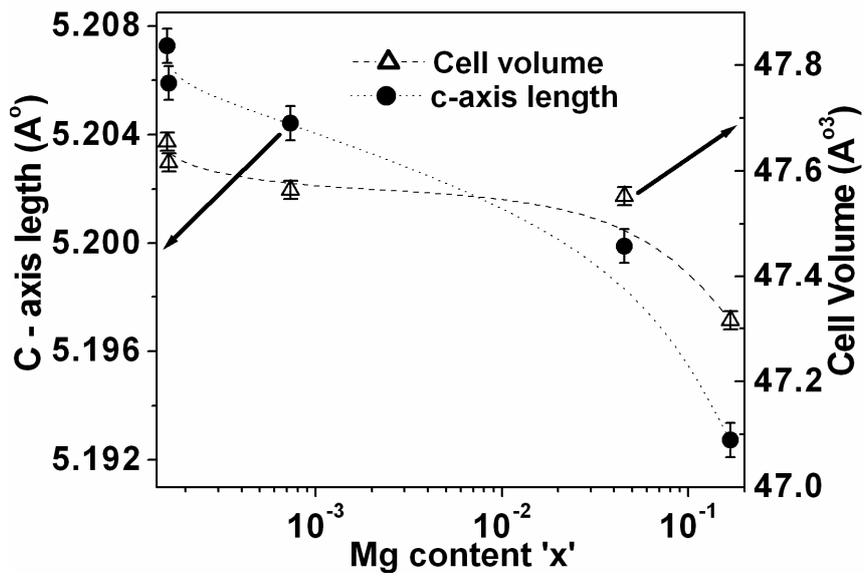

**Fig. 5(b)**

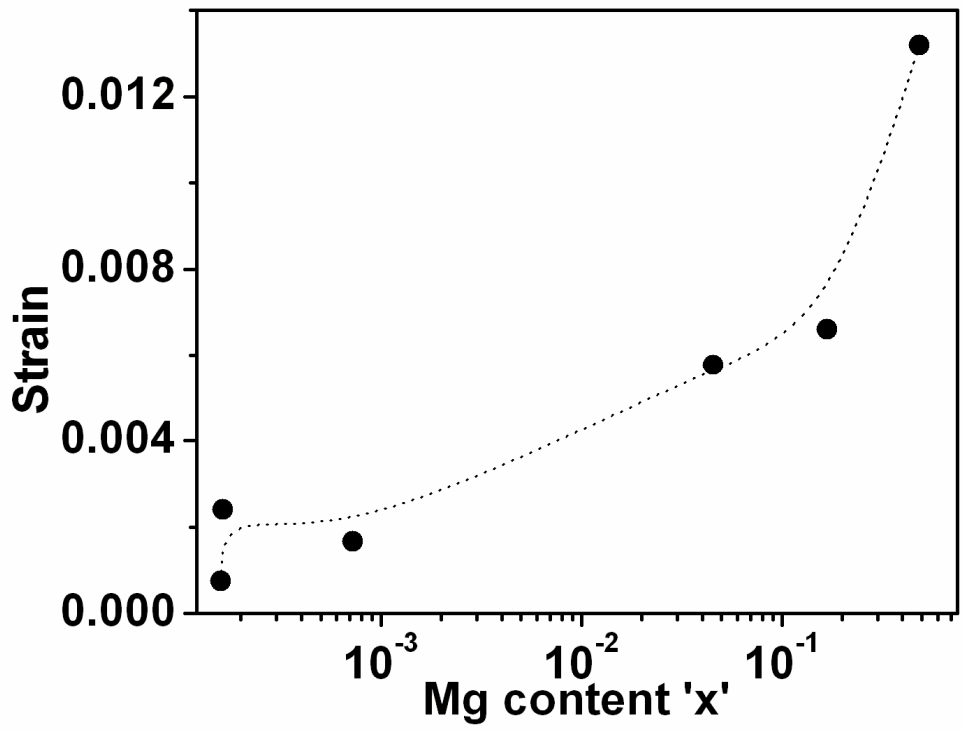

Fig. 6

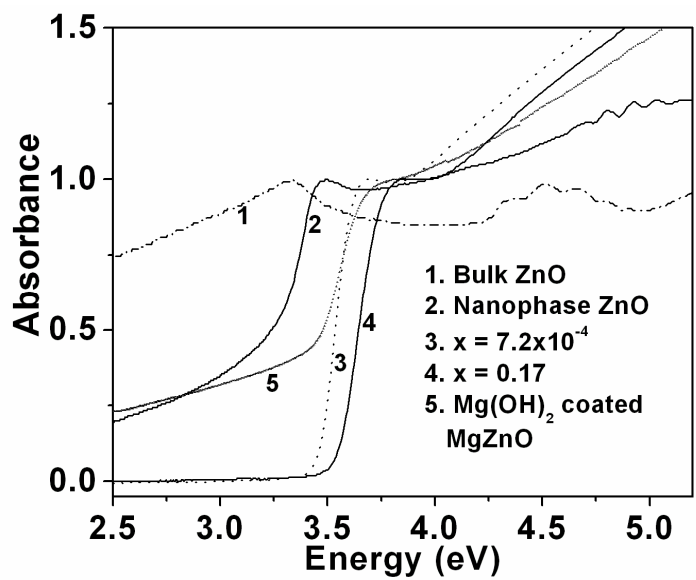

**Fig. 7(a)**

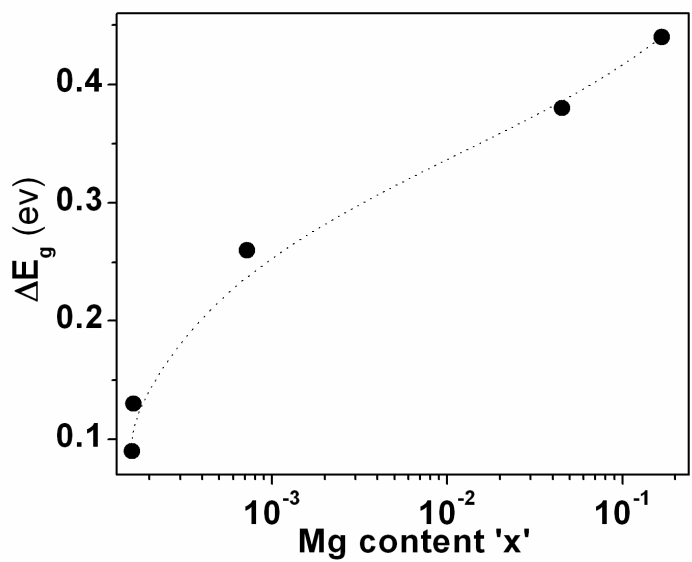

**Fig. 7(b)**

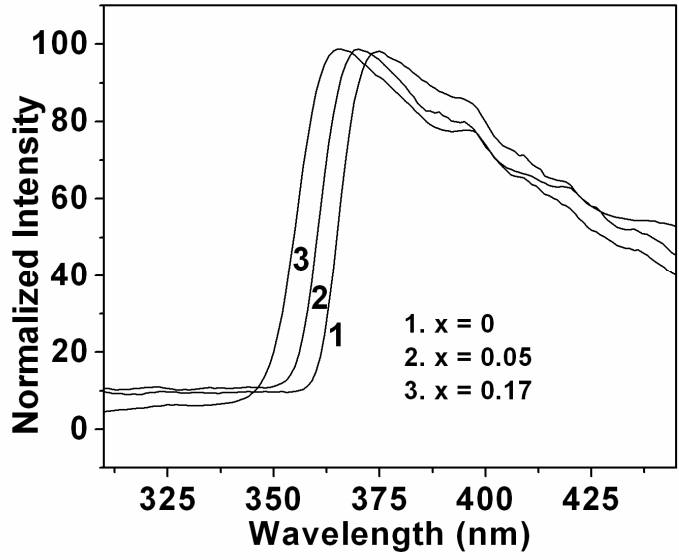

**Fig. 8(a)**

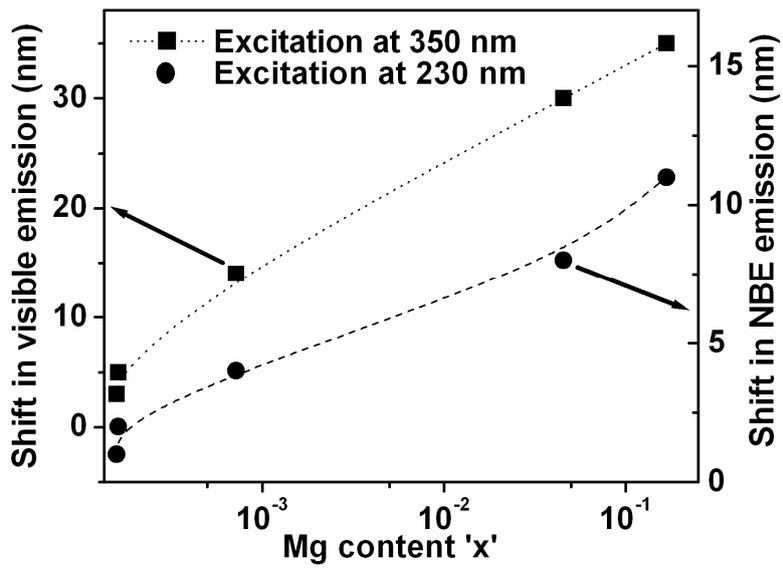

**Fig. 8(a)**

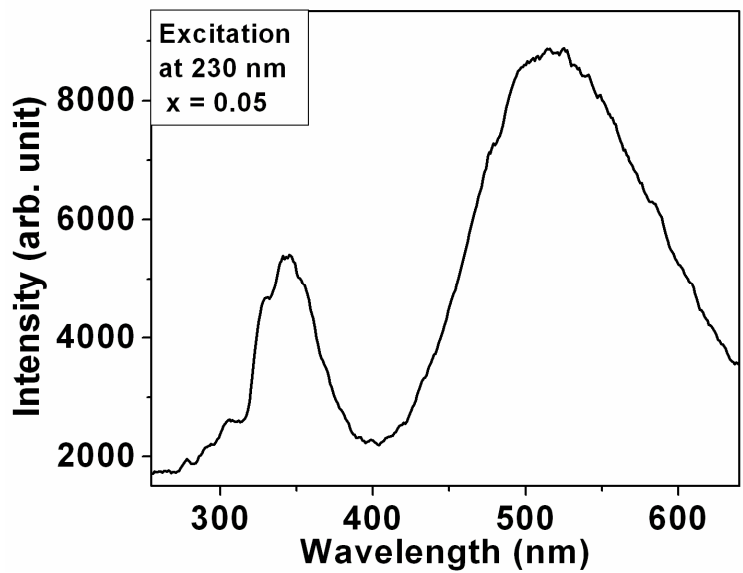

Fig. 9(a)

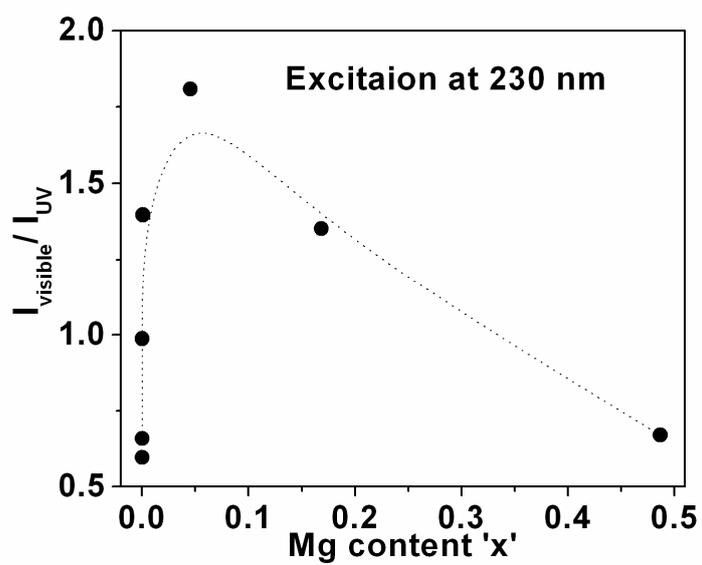

Fig. 9(b)